# Gravitational Back Reaction of Hawking Modes

## David Ring

2023 Clipper Dr, Lafayette, CO 80026
physicspapers@davering.com

**Abstract**
The radiation rate of an evaporating black hole is calculated in a toy model in which the geometry outside the collapsing matter is described by a Vaidya metric. When back reaction consistency is imposed, the singularity in the blueshift factor near the horizon is softened, suppressing the evaporation rate in the Schwarzschild case by the fourth power of the external time, thus rendering the hole eternal.

PACS number: 04.70.Dy

## 1. Introduction

It has been three decades since Hawking made the remarkable claim that black holes radiate thermally[1]. From the beginning, authors cautioned that the calculation should be done in spacetime consistent with the stress energy of the radiated matter[1,2]. Despite considerable interest, this calculation has never been performed in a realistic model of a black hole in 4 spacetime dimensions.

Section 2 describes qualitatively a geometry representing the standard radiation rate as calculated in the quasi-static geometry. Section 3 introduces back reaction in the single-particle case. The multiparticle case is calculated explicitly using a simple approximation in section 4. Section 5 explicitly confirms the consistency of the ansatz of section 4. Section 6 provides an alternative derivation which avoids the difficult analysis of section 4. Final notes appear in section 7.

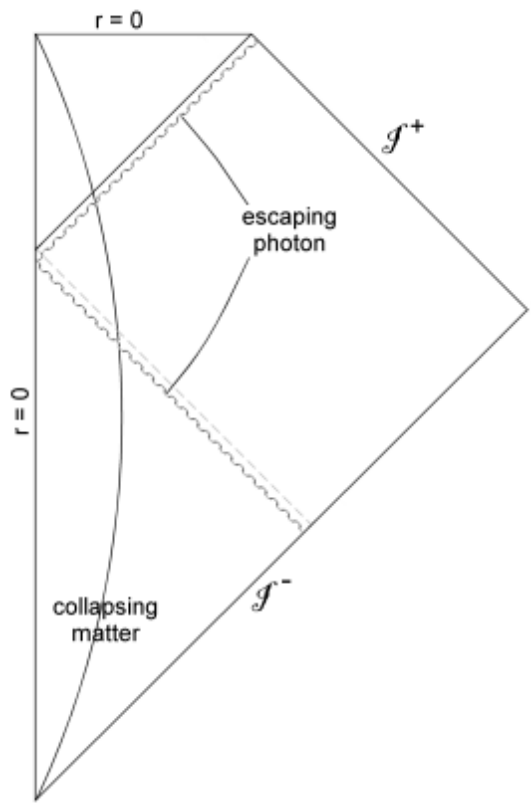 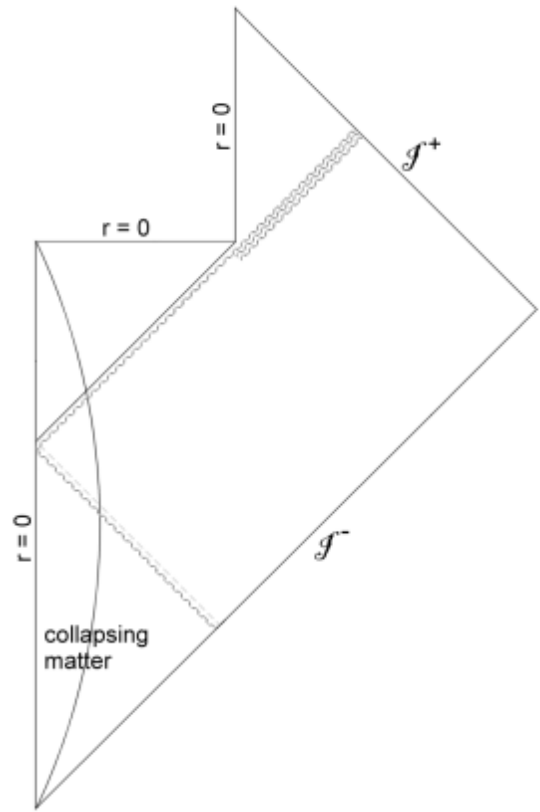

Fig. 1. Penrose diagram of a star collapsing to an eternal Schwarzschild hole. A mode just prior to the formation of the event horizon escapes to $\mathscr{I}^+$.

Fig. 2. The spacetime is extended to accommodate the disappearance of the hole following evaporation

## 2. Evaporation Geometry

Most calculations of the Hawking spectrum are performed in a static spacetime geometry (Fig. 1). The Bogoliubov coefficients $\beta_{ij}$ are nonzero, so the outgoing modes will be populated. Recognizing that the mass of the hole has radiated away, the back reaction is represented crudely by extending the final R=0 point to accommodate the future empty spacetime (Fig. 2).

To get a better understanding of the details of the mass reduction, consider a single localized (in retarded time) radial photon leaving a Schwarzschild hole. Setting aside the difficulty of setting up such a detector, one expects the spacetime to consist of two Schwarzschild geometries joined along the null surface of the photon (Fig. 3). The old apparent horizon at R = 2M is a null surface which now reaches $\mathscr{I}^+$. The new event horizon is located at R = 2 (M - δm), where δm is the energy of the measured photon.

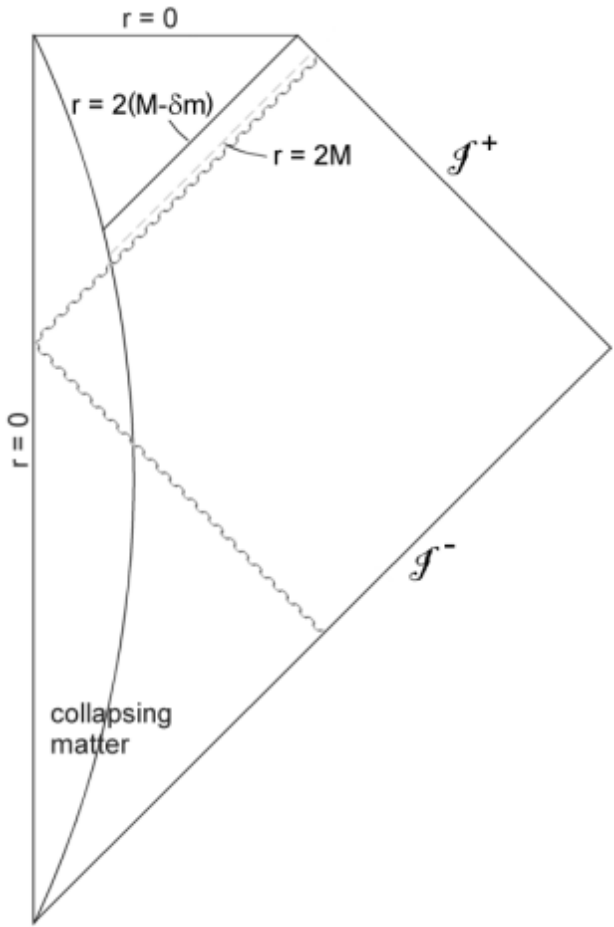
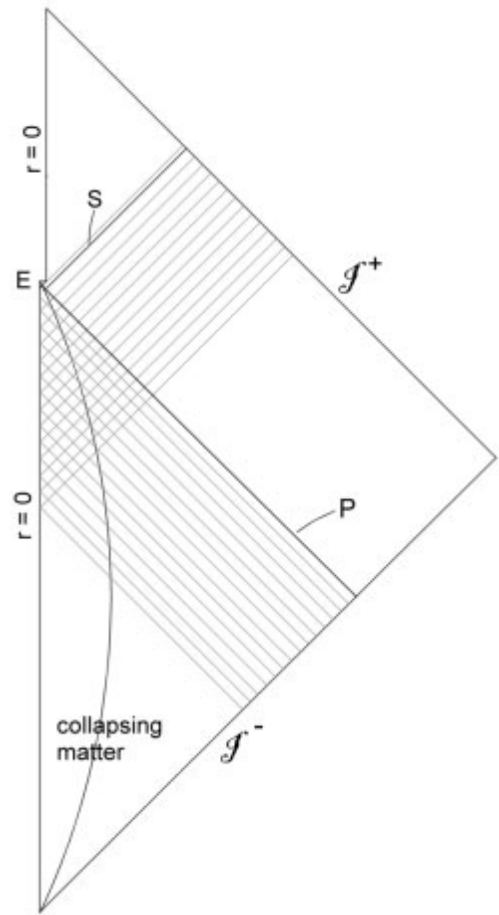

Fig. 3. The spacetime underneath an escaping photon is replaced with a Schwarzschild spacetime of smaller mass, giving a smaller apparent horizon.

Fig. 4. Each outgoing photon (gray line) reduces the hidden spacetime region until it has dimension $L_{pl}$. S is a null surface which leaves the hole just before the end-point explosion.

The spectrum of subsequent photons could be calculated in the 'quasi-static' geometry left behind by the first measured photon. Each photon changes the mass and temperature of the hole, and allows the previous apparent horizon to reach $\mathscr{I}^+$. Naively continuing this process until just before the hole reaches the Planck mass leads to Fig. 4. If we ignore the stress energy of the modes before they enter the collapsing star, as well as any virtual partners, and we smooth out the stress energy of the individual photons, the region outside the collapsing star can be described by a Vaidya metric[4]. This must be considered a toy model since the back reaction of the partners may counteract the effects of the escaping photons. The Vaidya metric is an exact solution of Einstein's equation describing a spherically symmetric null energy flux. Only qualitative features will be needed here.

Now consider a null spacetime surface (S in Fig. 4) just outside (prior to) the end-point explosion. Far from the hole, the surface surrounds ~$10^{16}$ erg of radiation. Close to the apparent horizon, the surface approaches R = 2 $L_{pl}$, where $L_{pl}$ is the Planck length. Continuing inward (backward in time) the surface must eventually intersect the infalling matter. That intersection must occur at R ≲ 2$L_{pl}$. This ensures the hidden spacetime region has dimensions of order $L_{pl}$. This argument extends in a natural way to massive particles, non-radial modes, and non-Schwarzschild holes.

## 3. Single-particle counterdistortion

When an escaping wavepacket is propagated back to $\mathcal{J}^-$ in the static geometry of Fig. 1, a negative frequency component arises due to the distortion of the packet near the event horizon. The blueshift is different across the width of the packet, and the phases in the trailing tail of the wave are more severely compressed than those in the leading tail (Fig. 5). The density of the phases (d$\varphi$/dv, where $\varphi$ is the phase of the wavepacket and v is the advanced time on $\mathcal{J}^-$) is inversely proportional to (v-$v_0$), where $v_0$ is the advanced time corresponding to the event horizon. The advanced and retarded times are given as usual by

(1) $$v = t + r + 2M \log\left|\frac{r}{2M} - 1\right|,$$

(2) $$u = t - r - 2M \log\left|\frac{r}{2M} - 1\right|.$$

Natural gravitational units where k = $\hbar$ = G = c = 1 are used throughout. The Hawking effect is demonstrated by calculating the negative frequency component of the distorted wave.

In the geometry of Fig. 3, the trailing tail propagates 'underneath' the majority of the stress energy of the packet and its compression should be measured relative to the underlying horizon. For any given frequency and packet width defined on $\mathcal{J}^+$, it is always possible to find a retarded time beyond which $v_L$-$v_0$<$v_T$-$v_1$ (see Fig. 6 for an explanation of the various v values). This counterdistortion has the potential to suppress the Bogoliubov coefficient for late photons. A complete analysis of this effect for a single particle is difficult, and it is likely that the wavepacket retains a negative frequency component due to the fact that the counterdistortion is not uniform.

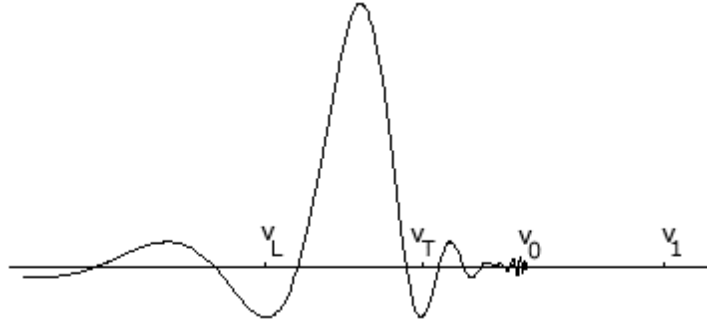

Fig. 5. A wavepacket having a symmetric form on $\mathcal{J}^+$ is distorted by uneven blueshift factors in the static Schwarzschild geometry. A null surface starting from $v_0$ will coincide with the event horizon. The local blueshift factor is inversely proportional to $v_0-v$. $v_L$ and $v_T$ are arbitrarily chosen to represent the leading and trailing parts of the wave respectively. Wavepackets which leave the hole at late times will have $v_L$ and $v_T$ arbitrarily close to $v_0$.

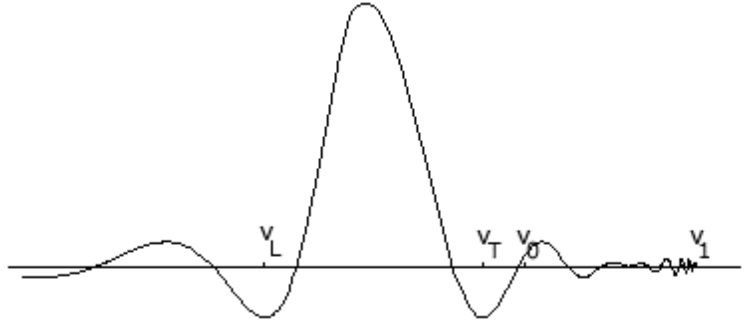

Fig. 6. The trailing part of the same wavepacket in the single-particle geometry of Fig. 3 receives a counterdistortion due to gravitational back reaction. A null surface starting from $v_1$ will coincide with the new event horizon. The local blueshift factor at $v_T$ is approximately inversely proportional to $v_1-v_T$, while at $v_L$ the factor remains approximately inversely proportional to $v_0-v_L$. For late photons, $v_0-v_L$ will be much smaller than $v_1-v_T$.

## 4. Multi-particle suppression

Some insight into the multiparticle case can be gained by considering a similar analysis in the smoothed version of the geometry of Fig. 4, a Vaidya metric representing continuous evaporation. If normal modes are defined on $\mathcal{J}^+$ and propagated backwards in time, the phases will be nearly uniformly distributed over $\mathcal{J}^-$ (compare to equation 2.17 of reference [1]). If a wavepacket is formed around a frequency $\omega$ on $\mathcal{J}^+$, it will exhibit negligible distortion across the width of the packet on $\mathcal{J}^-$ (compare to equation 4.14 of reference [2]). Indeed the frequencies will be sharply peaked around a large positive blueshifted frequency $\omega'$. The Bogoliubov coefficient will vanish to high orders.

This raises a paradox. Steady radiation eliminates the distortion, and leads to zero Bogoliubov coefficient. But if radiation is suppressed altogether, the static geometry is recovered, and the smooth distribution of phases has enough slope to restore the Bogoliubov coefficient.

The paradox is resolved if the radiation rate slows, but never reaches zero. To estimate the magnitude of the asymptotic radiation rate, assume the geometry is such that the blueshift factor for late wavepackets is given by

$$\frac{dv}{du} \sim a u^{-\alpha}, \quad \alpha > 1, \, a > 0 \tag{1}$$

where $\alpha$ and $a$ are constants. We will explicitly show that a finite number of particles are created in this geometry for any given frequency. We will then calculate the late time scaling behavior of the radiation using a simple approximation, and the back reaction of that radiation will be compared to (1) to solve for $\alpha$. Note that the static geometry has $dv/du = Ce^{-\kappa v}$, where $\kappa$ is the surface gravity and C is a constant. No exponential gives a consistent asymptotic result.

Following [1], the part of an outgoing normal mode of frequency $\omega$ that enters the collapsing body is given on $\mathcal{I}^-$ (for $v_0-v$ small and positive) by

$$p_\omega^{(2)} \sim \exp\left[i\omega\left(\frac{\alpha-1}{a}(v_0-v)\right)^{-1/(\alpha-1)}\right]. \tag{2}$$

In this calculation we are dropping factors which depend only on $\omega$. The Bogoliubov coefficient is found via fourier transform:

$$\alpha_{\omega\omega'}^{(2)} \sim \omega'^{1/2} \int_{-\infty}^{v_0} p_\omega^{(2)} e^{-i\omega' v} dv. \tag{3}$$

This can be evaluated for large $\omega'$ by the method of stationary phase to give

$$\alpha_{\omega\omega'}^{(2)} \sim \omega'^{\frac{1-3\alpha}{2\alpha}} e^{-i\omega' v_0} \exp\left[i\frac{\omega\alpha}{\alpha-1}\left(\frac{\omega' a}{\omega}\right)^{1/\alpha}\right]. \tag{4}$$

The $\beta$ coefficient is found by analytic continuation:

$$\beta_{\omega\omega'}^{(2)} \sim -i\alpha_{\omega(e^{i\pi}\omega')}^{(2)}$$

$$\sim \omega'^{\frac{1-3\alpha}{2\alpha}} e^{i\varphi} \exp\left[-\frac{\omega\alpha}{\alpha-1}\left(\frac{a\omega'}{\omega}\right)^{1/\alpha} \sin\frac{\pi}{\alpha}\right], \tag{5}$$

where $\varphi$ is a phase. The expectation value of the total number of created particles at $\mathcal{I}^+$ in the frequency range $\omega$ to $\omega+d\omega$ is $d\omega \int_0^\infty |\beta_{\omega\omega'}|^2 d\omega'$. Because of the exponential, this integral is finite for any $\alpha > 1$.

To find the time dependence of the radiation rate in this geometry, we need to find the magnitude of the negative frequency components for individual wavepackets at asymptotically late times. This can be estimated as a function of the distortion factor

(6) $$D = -\frac{d^2v/du^2}{dv/du} = \frac{\alpha}{u}.$$

Since this factor varies slowly over the width of a late wavepacket, the number expectation value of bosonic particles in a given mode approximates the expression in the static geometry, *ie.*

(7) $$N_{jn} = \Gamma_{jn}\left(\exp(2\pi\omega_j/D)-1\right)^{-1}$$
$$= \Gamma_{jn}\left(\exp\left[\frac{4\pi^2 n}{\alpha}(j+1/2)\right]-1\right)^{-1},$$

where the form of wavepackets of [1] has been assumed. The frequency $\omega_j$ is $(j+1/2)\epsilon$, and the packet is peaked around retarded time $u=2\pi n/\epsilon$ with width $1/\epsilon$, $\epsilon<<1$. $\Gamma_{jn}$ is the fraction of the $\mathcal{I}^+$ wavepacket which enters the collapsing body. Taking $n=1$ and $\epsilon=2\pi/T$ corresponds roughly to a physical situation where a detector surrounding the hole at r=4M is turned on at time t=4M+T(1-1/4$\pi$) and turned off at time t=4M+T(1+1/4$\pi$). At any given time scale T>>M, the radiation is dominated by the longest wavelength and earliest mode possible, *ie.* j=0. At such long wavelengths, modes which have non-zero angular momentum are scattered by the geometry, and $\Gamma_{jn} \to 0$. For zero angular momentum modes, $\Gamma_{jn}$ scales as the square of the frequency. Thus the evaporation rate scales as dM/dT $\propto$ T$^{-4}$ as there are ~T$^{-2}$ particles of energy $\epsilon/2$ leaving per time $2\pi/\epsilon$. Remarkably, this long term scaling behavior is valid for all $\alpha>1$, although the analysis is more involved when $\alpha$ is large. The correct value of $\alpha$ will be calculated in section 6. The long term scaling behavior is also robust despite corrections to (7) due to non-uniformity of D. The robustness is a result of the time dependence within the exponential. Wavepackets of any given frequency are cutoff exponentially within a few cycles, and our approximation becomes more accurate at the same time scales.

The total energy that escapes the final event horizon is likely to be ~1/M, and would be expected to occur for 0 < t < 4M as the radiation approaches the Hawking temperature. The total entropy of this radiation is O(1). The identification S = A/4 of the total entropy of the hole is dubious, but the differential version $\delta S = \delta A/4$ is still relevant for certain discussions. Certainly it would be interesting to study these effects in the presence of a thermal bath. Presumably the hole would cool until it reaches equilibrium, or gain mass indefinitely.

The scaling analysis following (7) does not apply in the case of rotating holes, where superradiant modes (see section 3 of [1]) dominate. Thus rotating holes must spin down before becoming truly black. If a hole has enough charge so that $m_e < e\Phi$, where $m_e$ is the electron mass, e is the electron charge, and $\Phi$ is the electrostatic potential on the horizon, then modes with $\omega < e\Phi$ will be fully populated. This is not superradiance since electrons are fermions, but it has the same overall effect here. The hole will lose charge before becoming black.

## 5. The back reaction

$\alpha$ can be calculated as follows. Consider two neighboring points along the surface of the infalling matter near the apparent horizon. Consider outgoing null geodesics passing through these points. The region outside the infalling matter and bounded by these two geodesics is approximately a slice of Schwarzschild geometry with mass M(*u*), where *u* is interpreted as the retarded time of the slice, rather than the coordinate within the slice. The surface of the infalling matter follows roughly an infalling null geodesic. The radius of this surface as a function of the retarded time *u* satisfies

$$(8) \quad \frac{dR}{du} = -\frac{R - 2M(u)}{2R}.$$

Taking M(*u*)-$M_\infty \to T^{-3}$, we find asymptotically R(*u*)-$2M_\infty \to u^{-3}$. For each outgoing null geodesic, the advanced time *v* on $\mathcal{J}^-$ is related to R(*u*) by d*v* = CdR, where C is a constant redshift factor which depends on the details of the collapse. (1) is recovered if $\alpha = 4$. Note that taking M constant recovers the static result d*v*/d*u* $\to Ce^{-\kappa v}$. If M decreases linearly, corresponding to continuing evaporation, d*v*/d*u* is roughly constant and wavepackets will receive negligible distortion.

## 6. Simple Calculation

These results can be derived in a way that avoids the difficult analysis of section 4. In the geometric optics approximation, the power radiated across a large sphere is given by [3]

$$(9) \quad \frac{dM}{dt} = \frac{\Gamma_0}{24\pi}\left[\frac{3}{2}\left(\frac{G''}{G'}\right)^2 - \frac{G'''}{G'}\right],$$

where G(u) = v relates coordinates on $\mathcal{J}^+$ and $\mathcal{J}^-$ via geometric ray tracing, and $\Gamma_0$ is the fraction of a zero angular momentum mode at $\mathcal{J}^+$ which enters the collapsing body. As above, most of the radiation has such long wavelengths that higher angular momentum modes are negligible.

As in section 5, taking M(u) constant in equation (8) gives G' = d*v*/d*u* $\to Ce^{-\kappa v}$, and dM/dt $\to -\Gamma_0/768\pi M^2$, in agreement with Hawking's result. Similarly, taking M = $M_0$-P*u* in equation (8) results in R $\to 2(M_0-Pu)(1+4P)$ and G' $\to -2PC(1+4P)$. Then dM/dt $\to 0$ to high order.

Taking M(u) = $M_\infty + au^{-\beta}$, we have R = $2M_\infty + 2au^{-\beta} + O(u^{-\beta-1})$. This gives

$$(10) \quad \frac{dM}{dt} = -\frac{\Gamma_0 u^{-2}}{48\pi}(\beta^2 - 1).$$

If we assume $\Gamma_0$ is constant, then $\beta$=1 and dM/dt vanishes. But late radiation is of much longer wavelength that the early radiation, and in fact $\Gamma_0 \sim u^{-2}$. Thus $\beta$=3, consistent with $\alpha$=4 from section 5.

## 7. Final Notes

It bears repeating that this geometry must be considered a toy model. There is a vacuum polarization term in the quantum matter distribution, whose back reaction tends to focus outgoing modes. This term may be interpreted as the stress energy of the infalling virtual partners of the outgoing particles. The intersection of the surface S in Fig. 4 with the infalling matter no longer need occur at $R \lesssim 2L_{pl}$. It may in fact occur near the original apparent horizon at R~2M. The vacuum polarization can restore the exponential singularity in the blueshift factor near the horizon, and thus the non-zero radiation rate. A more detailed calculation is in progress to determine if this is the case.